\documentclass[preprints,article,submit,oneauthor,pdftex]{mdpi}
\firstpage{1} 
\pdfoutput=1
\usepackage{marvosym}
\usepackage{amsmath}
\usepackage{amsfonts}
\usepackage{amssymb}
\usepackage{mathtools}

\Title{\:On physical optics approximation of viscous \\ \;stratified shear~flows}
\Author{\; Natanael Karjanto \orcidA{}}
\AuthorNames{Natanael Karjanto}
\address{Department of Mathematics, University College, Sungkyunkwan University, Natural Science Campus \\ 
2066 Seobu-ro, Jangan-gu, Suwon 16419, Gyeonggi-do, Republic of Korea \qquad \; \Letter: \url{natanael@skku.edu}}

\abstract{We study a mathematical model of a perturbed stratified shear mean flow in the presence of eddy coefficients of turbulent viscosity. We adopt the standard Boussinesq approximation in the natural convection of the buoyancy-driven flow and neglect the influence of the eddy coefficients of turbulent diffusivity. Comprising both the vertical and horizontal viscosity effects, a model for the vertical velocity perturbation corresponds to a fourth-order Taylor-Goldstein (TG) differential equation. Considering only the latter, we obtained a modified TG equation with the same order as the classical, inviscid one. Under an assumption of the slowly varying Brunt-Väisälä frequency and background horizontal velocity, we discuss the corresponding geometrical and physical optics approximations of the modified TG equation using the WKB method. We further investigate the behavior of these asymptotic solutions near singular values of a turning point and critical level.}

\keyword{stratified shear flow; eddy viscosity; modified Taylor-Goldstein equation; WKB method; geometrical optics approximation; physical optics approximation; matrix exponential; fundamental matrix; turning point; critical level} 

\begin{document}

\section{Introduction}

We consider dynamics of internal gravity waves in the atmosphere for continuously stratified parallel flows in the presence of horizontal eddy coefficients of turbulent viscosity. Under the Boussinesq approximation, we derive a vertical velocity perturbation equation from the two-dimensional Euler equations as the governing model. By including both the horizontal and vertical eddy viscosity, an equation for the vertical velocity perturbation is given by a fourth-order ordinary differential equation (ODE). On the other hand, in the absence of vertical eddy viscosity, it is modeled by a second-order ODE, an extended version of the classical Taylor-Goldstein (TG) equation for the inviscid case~\cite{taylor31,goldstein31,haurwitz31}. The latter governs the dynamics of Kelvin-Helmholtz (KH) instability, which in general refers to the instability for continuous distribution of velocity and density~\cite{kelvin71,helmholtz68}.

The study of flow with viscosity has a long history, starting from the Orr-Sommerfeld equation, an eigenvalue equation that models two-dimensional linear stability problem for nearly parallel viscous flows in a straight channel or a boundary layer~\cite{orr07a,orr07b,som08}. An extended version of the TG equation to include the effects of constant viscosity and diffusivity is modeled by Koppel using a sixth-order ODE~\cite{koppel64}. Hazel solved numerically the higher-order equation for the vertical velocity of internal gravity waves with kinematic viscosity and thermal conductivity~\cite{hazel67}. A couple of years later, the infinitesimal stability of inviscid, parallel, stratified shear flows described by the TG equation was studied numerically~\cite{hazel72}. For a mathematical perspective of instability in viscous fluids, see~\cite{renardy03}.

The TG equation and its variants have been used as mathematical models in various physical settings where stratified fluids or density stratification and shear flow occur. The interest lies in hydrodynamic stability since it sheds light on the problem of transition from the laminar to turbulent flow. Some examples where the TG equation aids in understanding this stability include surface gravity waves in the atmosphere~\cite{moninger76,janssen85a,janssen85b}, internal gravity waves in a stably stratified boundary layer~\cite{anne85,chimonas89}, and internal waves in the ocean~\cite{ssm98,watson94,thorpe99}, cf.~\cite{hall93,lawrence91,farrell93,sutherland05,zhou07,tedford09,camassa13,puetz18}. 

There exists, however, theoretical studies of the TG equation. For instance, Baldwin and Roberts studied the critical layer in a stratified shear flow where they introduced small viscosity and heat conductivity to remove the singularity~\cite{baldwin70}. Banerjee presented a geometry of the TG equation and some generalized results related to the stability~\cite{banerjee05}. Using a variational approach, Hirota and Morrison developed an efficient method for finding marginally stable (neutral mode) solutions of the TG equation~\cite{hirota15,hirota16}.

The interest in studying the TG model has been growing for the past two decades. Pratt and collaborators proposed an extended TG equation for modeling the stratified shear flow of arbitrary cross-section~\cite{pratt00,deng03}. While these authors also formulated the corresponding semicircle theorem, Subbiah and Ganesh sharpened it by deriving a semi-ellipse theorem for unstable modes. In addition to estimating their growth rate, they also obtained the upper and lower bounds for the neutral modes phase velocity~\cite{subbiah08,ganesh09}.

After the classical works on the TG equation with viscosity in the 1970s~\cite{hazel72,baldwin70}, it was not until the 1990s that the work starts to blossom again. Hall discussed the nonlinear instability of a boundary layer on a heated flat plate positioned in an incoming flow~\cite{hall92}. Dando examined the nonlinear evolution of inviscid G\"{o}rtler vortices in three-dimensional boundary layers~\cite{dando96}. Ng and Reid considered the stability problem in plane Poiseuille flow with the combined effects of stratification and viscosity~\cite{ng97}. Mureithi~et~al. investigated the effect of buoyancy on the upper-branch linear stability characteristics of an accelerating boundary layer flow in Tollmien-Schlichting waves~\cite{mureithi97}. See also~\cite{watson01,hogg01,hogg02,thorpe12,lawrence13,xiao15}.

Fast forward to the past decade, Smyth et al. investigated properties of shear instabilities with small-scale turbulence represented by vertical eddy viscosity and diffusivity coefficients~\cite{smn11}. Liu et al. examined an extended TG equation by including the effects of small-scale turbulence on the stability of stratified shear flows~\cite{lts12}. Thorpe et al. investigated the marginal stability of a stably stratified shear flow with ambient turbulence represented by vertical small eddy viscosity and diffusivity~\cite{tsl13}. In some circumstances, they found that viscosity amplifies instability. Li et al.\ further examined the effects of vertically varying eddy viscosity and diffusivity on the KH instability of a stratified shear flow~\cite{lst15}. Additionally, Khani and Waite measured horizontal and vertical eddy viscosity in stratified turbulence using high-resolution simulations~\cite{khani13}. Recently, Lian et al. performed numerical computations for the stability analysis of stratified and parallel shear flows with the effects of small-scale turbulence described by eddy viscosity and diffusivity~\cite{lian20}.

Our focus in this article is a modified TG equation with the presence of horizontal eddy viscosity but the absence of vertical eddy viscosity. Under the assumption of slowly varying background horizontal velocity and Brunt-V\"ais\"al\"a frequency, we propose an asymptotic solution for this modified TG equation using the WKB method. The technique is named after and developed by physicists Gregor Wentzel, Hendrik Anthony Kramers, and L\'{e}on Brillouin in 1926. They refined the ideas of George Green and Joseph Liouville~\cite{gill82,steele76}. Although mathematician Harold Jeffreys had developed a general method for approximating solutions of second-order linear ODEs in 1923, his contribution is often neglected since the trios WKB were not really aware of this earlier work. Indeed, a better name for this method should be coined as the JWKB or WKBJ approximation~\cite{bender99,baines95}. There are in fact at least 12 people who contributed to the development of WKB theory, including Liouville, Green, Horn, Rayleigh, Gans, Jeffrey, Wentzel, Kramers, Brillouin, Langer, Olver, and Meyer~\cite{hinch91}.

The WKB method has some limitations. The so-called physical optics WKB approximation turns singular and vanishes at a turning point and critical level, respectively. In the vicinity of the critical level, i.e., a level where the mean fluid velocity is equal to the horizontal phase velocity, an internal gravity wave flow is unstable~\cite{booker67,geller75,fritts76}. The behavior of both 2D and 3D internal gravity wave packets approaching a critical level has been examined both analytically and numerically~\cite{winters89,winters92,winters94}. Shear instability is responsible for maintaining deep-cycle turbulence due to the extraction of energy from the mean flow near the critical level through the interaction of mean shear and subgrid-scale stress~\cite{ssm98}. Note that although this reference refers to the latter as Reynolds stress, mean shear generally interacts with non-resolving fluxes, which are not necessarily Reynolds stress.

We organize this article as follows. The first part of Section~\ref{mathmodel} presents a derivation of mathematical modeling for vertical velocity perturbation with eddy coefficients of turbulent viscosity. Subsection~\ref{tagoeq} focuses on a modified TG equation in the presence of horizontal coefficient eddy viscosity only. Section~\ref{secwkb} derives both the geometrical and physical optics approximation using the WKB method. Section~\ref{secbeha} presents the solution near a turning point and critical level for different values of wind shears. Section~\ref{conclusion} concludes from our discussion.
 
\section{Mathematical modeling} \label{mathmodel}

\subsection{An extended Taylor-Goldstein equation}

Consider the Navier-Stokes momentum equation for an incompressible flow of the Newtonian fluid~\cite{batchelor00,white06,kundu15}:
\begin{equation}
\frac{\partial \tilde{\mathbf{u}}}{\partial t} + \left(\tilde{\mathbf{u}} \cdot \nabla \right) \tilde{\mathbf{u}} = -\frac{1}{\rho_0} \nabla \tilde{p} + \tilde{\theta} \mathbf{e}_z + \frac{\partial}{\partial x} \left(\tilde{a}_H \frac{\partial \tilde{\mathbf{u}}}{\partial x} \right) + \frac{\partial}{\partial z} \left(\tilde{a}_T \frac{\partial \tilde{\mathbf{u}}}{\partial z} \right).	\label{navstoeq}
\end{equation}
The fluid motion satisfies this equation and the continuity equation $\nabla \cdot \tilde{\mathbf{u}} = 0$.
For each quantity, we write $q = Q + \delta q'$, where $\delta \ll 1$ is a small parameter, $Q$ denotes a background state, undisturbed flow, taken to be steady of slowly varying and horizontally uniform, but varying in the vertical direction, of the corresponding boundary layer; and $\delta q'$ is the first-order perturbation. Assume that the background flow is in the hydrostatic balance.

For another small parameter $\eta \ll 1$,  $\tilde{\theta} = \eta g\tilde{\theta}'/\Theta$ is the potential temperature. Furthermore, $g$ is the gravitational acceleration, $\tilde{p}$ is the pressure, $\rho_0$ is the reference density, $\tilde{a}_H(x,z,t)$ and $\tilde{a}_T(x,z,t)$ are the horizontal and the vertical eddy coefficients of turbulent viscosity, respectively. The continuity equation for potential temperature is given by
\begin{equation}
\frac{\partial \tilde{\theta}}{\partial t} + \left(\tilde{\mathbf{u}} \cdot \nabla \right) \tilde{\theta} = 
\frac{\partial}{\partial x} \left(\tilde{k}_H \frac{\partial \tilde{\theta}}{\partial x} \right) + 
\frac{\partial}{\partial z} \left(\tilde{k}_T \frac{\partial \tilde{\theta}}{\partial z} \right), \label{conteqpottem}
\end{equation}
where $\tilde{k}_H(x,z,t)$ and $\tilde{k}_T(x,z,t)$ are the horizontal and the vertical eddy coefficients of turbulent diffusivity, respectively. In what follows, we will assume that $\tilde{a}_H$, $\tilde{a}_T$, $\tilde{k}_H$, and $\tilde{k}_T$ are independent of $x$, following the argument in~\cite{lts12,tsl13,lst15}, cf.~\cite{kundu15,garratt92,kantha00}. In general, however, they should depend on $(x,z,t)$, as shown recently that even at anisotropic grid points with $x \gg z$, both eddies in horizontal and vertical directions are in fact coupled~\cite{khani20}. 

The Navier-Stokes equation~\eqref{navstoeq} and the continuity equation for potential temperature~\eqref{conteqpottem} admit the initial and boundary conditions $\tilde{\mathbf{u}}(x,z,0) = \tilde{\mathbf{u}}_0(x,z)$, and $\tilde{\mathbf{u}}(0,z,t) = \mathbf{f}_1(z,t)$, $\tilde{\mathbf{u}}(x,0,t) = \mathbf{f}_2(x,t)$, respectively. They also possess an exact stationary solution for the flow velocity $\tilde{\mathbf{u}} = (\tilde{u},\tilde{v},\tilde{w})$, potential temperature, and pressure, where the former is written in the component form as follows:
\begin{equation*}
\tilde{u} = U(z), \quad \quad \tilde{v} = 0, \quad \quad \tilde{w} = 0, \quad \quad \tilde{\theta} = \Theta(z), \quad \quad \text{and} \quad \quad \tilde{p} = P(z).
\end{equation*}
Here, $U(z)$, $\Theta(z)$, and $P(z)$ denote the horizontal mean flow, potential temperature, and pressure at the basic state, respectively. This indicates that the equilibrium of a stratified fluid in gravity field can be achieved when these quantities depend only on the vertical height~$z$. Hence, the flow velocity vector with first-order perturbation can be written as $\tilde{\mathbf{u}} = (U(z) + \delta u, 0, \delta w)$.

The linearized equations in the presence of eddy viscosity and diffusivity terms read
\begin{align}
\left(\frac{\partial }{\partial t} + U \frac{\partial}{\partial x}\right) \theta + w \frac{d \Theta}{dz} &= 
K_H \frac{\partial^2 \theta}{\partial x^2} + \frac{\partial}{\partial z} \left(K_T \frac{\partial \theta}{\partial z} + k_T \frac{d \Theta}{dz} \right), \label{eddythetaper}\\
\left(\frac{\partial }{\partial t} + U \frac{\partial}{\partial x}\right) u + w \frac{dU}{dz} &= 
- \frac{1}{\rho_0} \frac{\partial p}{\partial x} + A_H \frac{\partial^2 u}{\partial x^2} + \frac{\partial}{\partial z} \left(A_T \frac{\partial u}{\partial z} + a_T \frac{dU}{dz} \right),  \label{eddyuper} \\
\left(\frac{\partial}{\partial t} + U \frac{\partial}{\partial x}\right) w &= 
- \frac{1}{\rho_0} \frac{\partial p}{\partial z} + g \frac{\theta}{\Theta} + A_H \frac{\partial^2 w}{\partial x^2} + \frac{\partial}{\partial z} \left(A_T \frac{\partial w}{\partial z} \right), \label{eddywper}\\
\frac{\partial u}{\partial x} + \frac{\partial w}{\partial z} 	   &= 0. \label{eddyvoper}
\end{align}
We can exclude the pressure terms by subtracting $\partial_z$ of~\eqref{eddyuper} from $\partial_x$ of~\eqref{eddywper}. Applying a derivative operator $\left(\frac{\partial }{\partial t} + U \frac{\partial}{\partial x}\right) \frac{\partial}{\partial x}$ to this equation, and employing~\eqref{eddythetaper} and~\eqref{eddyvoper}, we can further eliminate $u$ and $\theta$ to obtain a single equation for $w$. In the absence of diffusivity, it is given as follows:
\begin{align}
\left(\frac{\partial }{\partial t} + U \frac{\partial}{\partial x}\right)^2 \nabla^2 w & + N^2 \frac{\partial^2 w}{\partial x^2} - \frac{d^2 U}{dz^2} \left(\frac{\partial}{\partial t} + U \frac{\partial}{\partial x}\right) \frac{\partial w}{\partial x} \nonumber \\
&= \left(\frac{\partial}{\partial t} + U \frac{\partial}{\partial x}\right) 
\left[\frac{\partial}{\partial z} \left(A_N \frac{\partial^3 w}{\partial z \partial x^2} \right) + A_H \frac{\partial^4 w}{\partial x^4} +
\frac{\partial^2}{\partial z^2} \left(A_T \frac{\partial^2 w}{\partial z^2} \right) \right], \label{singeq4w}
\end{align}
where $\nabla^2 = \partial_x^2 + \partial_z^2$ is a Laplacian operator, $N^2 = g \frac{d}{dz} \ln \Theta$ is the square of the  Brunt-V\"ais\"al\"a (buoyancy) frequency, and $A_N = A_H + A_T$~\cite{holton04,nappo13}. Substituting $w(x,z,t) = \hat{w}(z) e^{ik(x - ct)}$ into~\eqref{singeq4w}, we obtain a fourth-order ODE in $\hat{w}$, where the prime denotes differentiation with respect to~$z$:
\begin{equation}
\hat{w}'' + \left[\frac{N^2}{(U - c)^2} - \frac{U''}{U - c} - k^2 \right] \hat{w} = 
\frac{1}{ik(U - c)} \left[\frac{d^2}{dz^2} \left(A_T \hat{w}''\right) - k^2 \frac{d}{dz} \left(A \hat{w}' \right) + k^4 A_H \hat{w} \right], \label{4thorderode}
\end{equation}
with boundary conditions $\hat{w}(0) = \hat{w}_0$ and $\hat{w}'(0) = \hat{w}_1$. Note that the absence of the viscosity terms on the right-hand side of~\eqref{4thorderode} leads to the classical Taylor-Goldstein equation, which governs the behavior of a perturbed vertical velocity in continuously stratified parallel flows~\cite{taylor31,goldstein31}. For unidirectional flow, the Miles-Howard semicircle theorem is established~\cite{miles61,howard61}. It states that the complex-valued wave velocity $c$ of any unstable mode of a perturbed inviscid fluid parallel flows must lie inside the semicircle in the upper half of the $c$-plane with a diameter the range of $U$~\cite{huppert73,engevik85,miles86,bayly88}.

In the absence of stratification, Rayleigh's inflection-point theorem for inviscid, bounded, parallel, homogeneous shear flow is established. This necessary but not sufficient condition states that the flow may be linearly unstable to perturbations only if $U$ has an inflection point~$z_{\ast}$~\cite{rayleigh80}. Furthermore, a stronger condition for instability was obtained by Fj{\o}rtoft, who showed that $U''(U - U_s) < 0$ somewhere in the field flow, where $z_{\ast}$ is an inflection point at which $U''(z_{\ast}) = 0$ and $U_s = U(z_{\ast})$~\cite{fjortoft50}, see also~\cite{squire33,hoiland53,drazin02,drazin04}, cf.~\cite{gubarev07,gubarev13} for further discussion on necessary and sufficient conditions. In what follows, we will consider the special case where $A_T = 0$ and nonzero $A_H$.

\subsection{Modified Taylor-Goldstein equation with horizontal eddy viscosity} \label{tagoeq} 

Let $U(z)$ be the background velocity, $A_H(z)$ be the horizontal eddy viscosity, where both are real-valued functions of the vertical coordinate $z$. Let $c = c_r + i c_i \in \mathbb{C}$ with $c_i > 0$ for unstable situation. We write $U(z) - c = u_1(z) = \rho_1 e^{i \phi_1}$ and $U(z) - c - ikA_H = u_2(z) = \rho_2 e^{i \phi_2}$, where both the amplitudes $\rho_{1,2}(z)$ and the phases $\phi_{1,2}(z)$ are real-valued functions. The square of the Brunt-V\"ais\"al\"a frequency is defined as $N^2 = g \frac{d}{dz} \ln \Theta(z)$, with $\Theta$ as the basic state of potential temperature. Consider a modified TG equation for the vertical velocity profile $\hat{w}$ in the absence of $A_T$ but only the presence of $A_H$. It reads
\begin{equation}
\hat{w}'' + Q_1(z) \hat{w}' + Q_0(z) \hat{w} = 0, \label{tago1}
\end{equation}
where both $Q_0$, $Q_1 \in \mathbb{C}$ and are given as follows:
\begin{align}
Q_0(z) &= \frac{1}{u_2} \left(\frac{N^2}{u_1} - U'' \right), \\
Q_1(z) &= -ik\frac{A_H}{u_2}.
\end{align}
Applying the following Liouville transformation by introducing a new variable $\tilde{w}$ and an integration factor 
\begin{equation}
\hat{w}(z) = \text{exp}\left(-\frac{1}{2} \int^z Q_1(\zeta) \, d\zeta \right) \tilde{w}(z), 		\label{liouv}
\end{equation}
we can simplify the modified TG equation~\eqref{tago1} by removing the first-order derivative term. It yields
\begin{equation}
\tilde{w}'' + Q_2(z) \tilde{w} = 0, \label{tago2}
\end{equation}
where 
\begin{equation}
Q_2(z) =  Q_0 - \frac{Q_1^2}{4} - \frac{Q_1'}{2}.
\end{equation}

Rescaling the vertical coordinate $z$ as $Z = \varepsilon z$, where $\varepsilon \ll 1$ is a small dimensionless parameter, then the modified TG equation~\eqref{tago2} reads
\begin{equation}
\varepsilon^2 \ddot{\tilde{w}} + Q_\varepsilon(Z) \tilde{w} = 0, \label{tago3}
\end{equation}
where the complex-valued potential term $Q_\varepsilon$ is given by
\begin{equation}
Q_\varepsilon(Z) = \frac{N^2}{u_1 u_2} - k^2 - \frac{\varepsilon^2}{u_2} \left(\frac{u_3}{u_2} + u_4 \right), \label{Qeps} 
\end{equation}
with
\begin{align*}
u_3 &= \frac{1}{4}k^2 (\dot{A_H})^2 + \frac{1}{2}ik \dot{U} \dot{A_H}, \\ 
u_4 &= \ddot{U} - \frac{1}{2}ik \ddot{A_H}.
\end{align*}
Note that the differentiation with respect to the scaled vertical variable $Z$ has now been replaced by a dot.

\section{WKB method and matrix exponential}  \label{secwkb}

WKB theory provides relatively simple approximate solutions for the Fourier component of the perturbed vertical velocity $\tilde{w}$, which is expressed as a linear second-order ODE given by the modified TG equation~\eqref{tago2} or~\eqref{tago3}. When the frequency-like term $Q_\varepsilon$~\eqref{Qeps} varies slowly vertically, an approximate solution to~\eqref{tago3} can be sought by implementing the WKB method. 

WKB theory is a particular case of multiple-scale analysis and an early application of this method comes from quantum mechanics when the conditions of the particle motion are nearly classical. This method does not only enjoy applications but also has been successfully implemented in the lower troposphere and boundary layer regions, including internal gravity~\cite{gill82,breth66,pedlosky03}, atmospheric acoustic-gravity~\cite{pitt65,einau70}, radio~\cite{budden88}, and mountain waves~\cite{laprise93,griso94,teix04a,teix04b,teix06,miran09,teix09}.

\subsection{WKB approximation}

Consider the modified TG equation~\eqref{tago3} and write its potential term $Q_\varepsilon$ by separating the real and imaginary parts, given as follows:
\begin{equation*}
Q_\varepsilon(Z) = A(Z) + i B(Z) - \varepsilon^2 \left[C(Z) + i D(Z) \right]
\end{equation*}
where
\begin{align*}
A(Z) &= \frac{N^2 \left[(U - c_r)^2 - c_i^2 - k c_i A_H \right]}{\left[(U - c_r)^2 - c_i^2 - k c_i A_H \right]^2 + (U - c_r)^2 (2c_i + kA_H)^2} - k^2 \\
B(Z) &= \frac{N^2 (U - c_r) (2c_i + kA_H)}{\left[(U - c_r)^2 - c_i^2 - kc_i A_H \right]^2 + (U - c_r)^2 (2c_i + kA_H)^2} \\
\tilde{C}(Z) &= \frac{1}{4} k^2 \left(\dot{A_H}\right)^2 \left[(U - c_r)^2 - c_i^2\right] - k c_i(U - c_r) \dot{U} \dot{A_H} 
+ \ddot{U} (U - c_r) \left[(U - c_r)^2 + c_i^2 + 2 k c_i A_H \right] \\ 
& \qquad + \frac{1}{2} \ddot{A_H} \left[k(U - c_r)^2 \left(2 c_i -  k A_H\right) + k^2  c_i^2 A_H \right] \\
C(Z) &= \frac{\tilde{C}(Z)}{\left[(U - c_r)^2 - c_i \right]^2 + 4 c_i^2 (U - c_r)^2} \\
\tilde{D}(Z) &= 2 c_i \ddot{U} (U - c_r)^2 + \frac{1}{2} k^2 c_i (U - c_r) \left[\left(\dot{A_H}\right)^2 - 2A_H \ddot{A_H} \right] \\ 
& \qquad - \frac{1}{2} \left[(U - c_r)^2 - c_i^2 \right] \left[2 \ddot{U} (c_i + kA_H) - k \dot{U} \dot{A_H} + k (U - c_r) \ddot{A_H} \right] \\
D(Z) &= \frac{\tilde{D}(Z)}{\left[(U - c_r)^2 - c_i \right]^2 + 4 c_i^2 (U - c_r)^2}.
\end{align*}
Let also $M(Z) = m(Z) + i n(Z)$, $m, n \in \mathbb{R}$, where all three quantities $M$, $m$, and $n$ are expanded in the order of $\varepsilon$:
\begin{align*}
M(Z) &= M_0(Z) + \varepsilon M_1(Z) + \varepsilon^2 M_2(Z) + \dots  \\
     &= m_0(Z) + i n_0(Z) + \varepsilon \left[m_1(Z) + i n_1(Z) \right] + \varepsilon^2 \left[m_2(Z) + i n_2(Z) \right] + \dots. 
\end{align*}
Note that two conditions must be satisfied for the WKB approach to be valid and useful. First, the series $\varepsilon^{-1} \int M(Z)$ must be an asymptotic series in $\varepsilon$ as $\varepsilon \rightarrow 0$ uniformly for all $Z$ in the interval of approximation. Second, for the WKB solution truncated at the term $\varepsilon^{j - 1} M_{j}$, the next term must be small compared to one for all $Z$ in the interval of approximation, i.e., \ $\varepsilon^{j} |M_{j + 1}(Z)| \ll 1$ as $\varepsilon \rightarrow 0$~\cite{bender99}.

The WKB anzats for~\eqref{tago3} is given by
\begin{equation}
\tilde{w}(Z) = \tilde{w}(Z = 0) \, \text{exp} \left(\frac{1}{\varepsilon} \int_{0}^{Z} M(\zeta) \, d\zeta \right).
\end{equation}
Substituting this anzats to~\eqref{tago3}, collecting the terms in the order of $\varepsilon$, we obtain coupled equations by separating the real and imaginary parts. The lowest order equations are given by
\begin{align*}
m_0^2 - n_0^2 + A &= 0, \\
2 \, m_0 \, n_0 + B	  &= 0.
\end{align*}
Solving these equations, we obtain
\begin{align*}
m_0^2(Z) &=  \frac{1}{2} \left(-A + \sqrt{A^2 + B^2} \right)\\
n_0^2(Z) &=  \frac{1}{2} \left( A + \sqrt{A^2 + B^2} \right).
\end{align*}
By noting $m_0$ and $n_0$ as the positive roots of the expressions above, there are two fundamental solutions for the geometrical optics WKB approximation of~\eqref{tago3}:
\begin{equation*}
\text{exp} \left\{ \frac{1}{\varepsilon} \int_0^Z \left[m_0(\zeta) + i n_0(\zeta) \right] \, d\zeta \right\} \qquad \qquad \text{and} \qquad \qquad
\text{exp} \left\{-\frac{1}{\varepsilon} \int_0^Z \left[m_0(\zeta) + i n_0(\zeta) \right] \, d\zeta \right\}.
\end{equation*}

The corresponding solution represents a plane-wave with amplitude independent of the vertical variable $z$. This geometrical optics approximation satisfies 
\begin{equation*}
\tilde{w}'' + \left[(A + iB) \mp (m_0' + in_0') \right] \tilde{w} = 0,
\end{equation*}
where $A + i B + M_0^2 = 0$, which contributes from the lowest-order term of $Q_\varepsilon$. From here, we observe that for the WKB approximation to be valid, then we require that the condition $\left| M_0'(z) \right| \ll \left| M_0^2(z) \right|$ needs to be satisfied. Hence, we can obtain a reasonable definition for~$\varepsilon > 0$~\cite{teix04a}:
\begin{equation*}
\varepsilon = \max_{z \ \geq \ 0} \left| \frac{M'(z)}{M^2(z)} \right| \approx \max_{z \ \geq \ 0} \left| \frac{M_0'(z)}{M_0^2(z)} \right| = 
\max_{z \ \geq \ 0} \sqrt{\frac{\left(m_0'\right)^2 + \left(n_0'\right)^2}{A^2 + B^2}}.
\end{equation*}
As an example, for $U(z) = 1 + 2\tanh(z)$, $N^2(z) = 1 + 4\tanh^2(z)$, $c = 5 + 0.1 i$, $k = 0.1$, and $A_H(z) = 2e^{-z^2}$, any value of $\varepsilon$ satisfying $0 < \varepsilon \leq 0.3$ would be sufficient.

The first-order of $\varepsilon$ yields the following system of equations:
\begin{align*}
2 (m_0 m_1 - n_0 n_1) + m_0' &= 0, \\
2 (m_0 n_1 + m_1 n_0) + n_0' &= 0.
\end{align*}
Solving these equations and using the fact that $|M_0|^2 = m_0^2 + n_0^2 = \sqrt{A^2 + B^2}$, we obtain $m_1$ and $n_1$ and their integrals:
\begin{align*}
m_1(Z) &= -\frac{1}{4} \frac{\frac{d}{dZ}(m_0^2 + n_0^2)}{m_0^2 + n_0^2}, \\
\int_{0}^{Z} m_1(\zeta) \, d\zeta &=  \ln \sqrt{\left|\frac{M_0(0)}{M_0(Z)} \right|}, \\ 
n_1(Z) &= \frac{\frac{d}{dZ}(m_0/n_0)}{2 \left[1 + (m_0/n_0)^2 \right]}, \\
\int_{0}^{Z} n_1(\zeta) \, d\zeta &= \frac{1}{2} \tan^{-1} \left(\frac{m_0(Z)}{n_0(Z)} \right) - \frac{1}{2} \tan^{-1} \left(\frac{m_0(0)}{n_0(0)} \right).
\end{align*}
The physical optics approximation of the WKB method reads
\begin{equation}
\tilde{w}(z) = \tilde{w}(z = 0) \sqrt{\left|\frac{M_0(z = 0)}{M_0(z)} \right|} \, \text{exp} \left[\frac{1}{2} i (\alpha - \alpha_0) + \int_{0}^{z} M_0(\zeta) \, d\zeta \right],  \label{phyopt}
\end{equation}
where we have defined 
\begin{equation*}
\tan \alpha(z) = \frac{m_0}{n_0} = \pm \sqrt{1 + 2 \left(\frac{A}{B}\right)^2 - 2 \frac{A}{B}\sqrt{1 + \left(\frac{A}{B}\right)^2} }.
\end{equation*}
In the literature, what is called as the ``WKB solution'' usually refers to the physical optics approximation. The geometric approximation is rarely used when we discuss the WKB method, as it shortfalls to provide the correction for the spatially varying coefficients. 

The system of equations for the second-order $\varepsilon$ is given by
\begin{align*}
m_1^2 - n_1^2 + 2(m_0 m_2 - n_0 n_2) + m_1' &= C, \\
2\left(m_0 n_2 + m_1 n_1 + m_2 n_0 \right) + n_1' &= D.
\end{align*}
These solve as
\begin{align*}
m_2(Z) &= \frac{\left[C - m_1' - (m_1^2 - n_1^2) \right]m_0 + \left[D - n_1' - 2m_1 n_1 \right] n_0 }{2 \left(m_0^2 + n_0^2 \right)}, \\
n_2(Z) &= \frac{\left[D - n_1' - 2m_1 n_1 \right]m_0 - \left[C - m_1' - (m_1^2 - n_1^2) \right] n_0 }{2 \left(m_0^2 + n_0^2 \right)}.
\end{align*}

\begin{figure}[H]
\begin{center}
\includegraphics[width=0.35\textwidth]{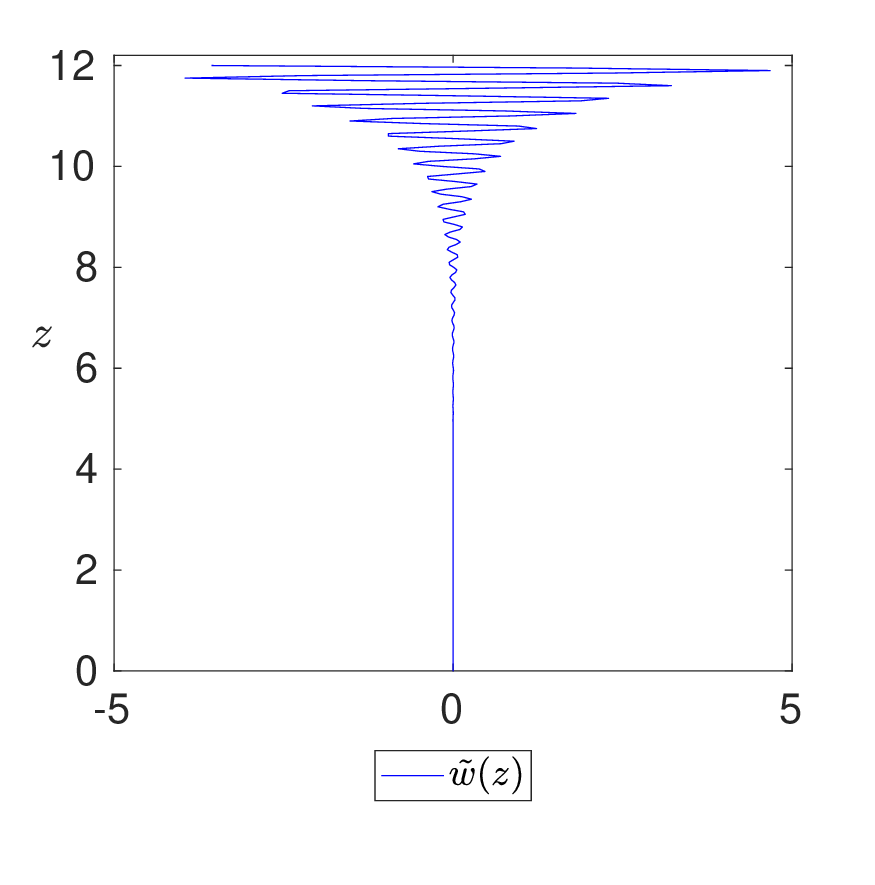} \hspace*{0.25cm}
\includegraphics[width=0.35\textwidth]{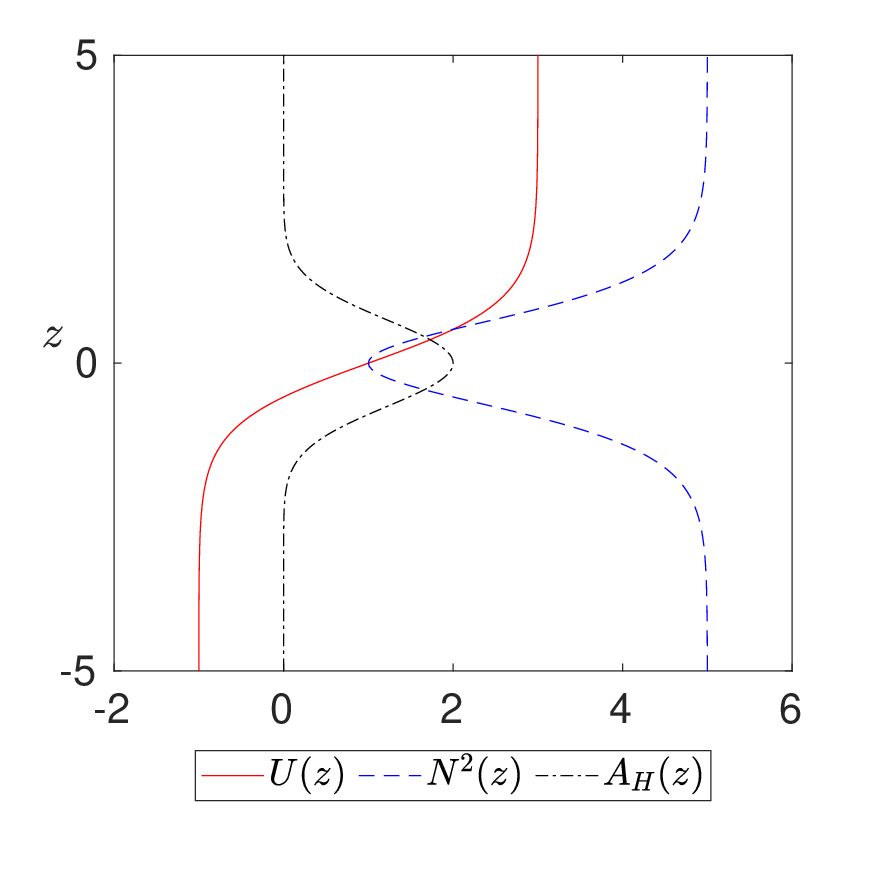} \vspace*{-0.5cm}
\end{center}
\caption{(Left panel) A plot of the WKB solution~\eqref{phyopt} at the onset of instability for a Gaussian profile of eddy viscosity. We have taken $U(z) = 1 + 2\tanh(z)$, $N^2(z) = 1 + 4\tanh^2(z)$, $c = 5 + 0.1 i$, $k = 0.1$, and $A_H(z) = 2e^{-z^2}$. There exists no critical level for this particular choice of parameters. (Right panel) The plots of $U$ (red), $N^2$ (dashed blue), and $A_H$ (dash-dotted black), in which all are symmetric with respect to $z = 0$. In both cases, the vertical axis is $z$.} \label{jwkbex}
\end{figure}

Although in principal we can proceed further to obtain an expression for the WKB approximation at higher-order of $\varepsilon$, integrating $M_2(Z)$ is already cumbersome at this stage due to its dependence on $C(Z)$ and $D(Z)$. Hence, the WKB approximation~\eqref{phyopt} suffices for our analysis. The left panel of Figure~\ref{jwkbex} displays a physical optics approximation of the WKB solution~\eqref{phyopt} for a Gaussian profile of horizontal eddy viscosity $A_H$ at the onset of instability. The horizontal background velocity and the square of the Brunt-V\"{a}is\"{a}l\"{a} frequency are taken as $U(z) = 1 + 2\tanh(z)$ and $N^2(z) = 1 + 4\tanh^2(z)$, respectively, which are shown in the right panel of Figure~\ref{jwkbex}. Both quantities are symmetric with respect to the $z$-axis.

\subsection{Matrix exponential and fundamental matrix physical optics approximations}

\begin{figure}[h!]
\begin{center}
\includegraphics[width=0.35\textwidth]{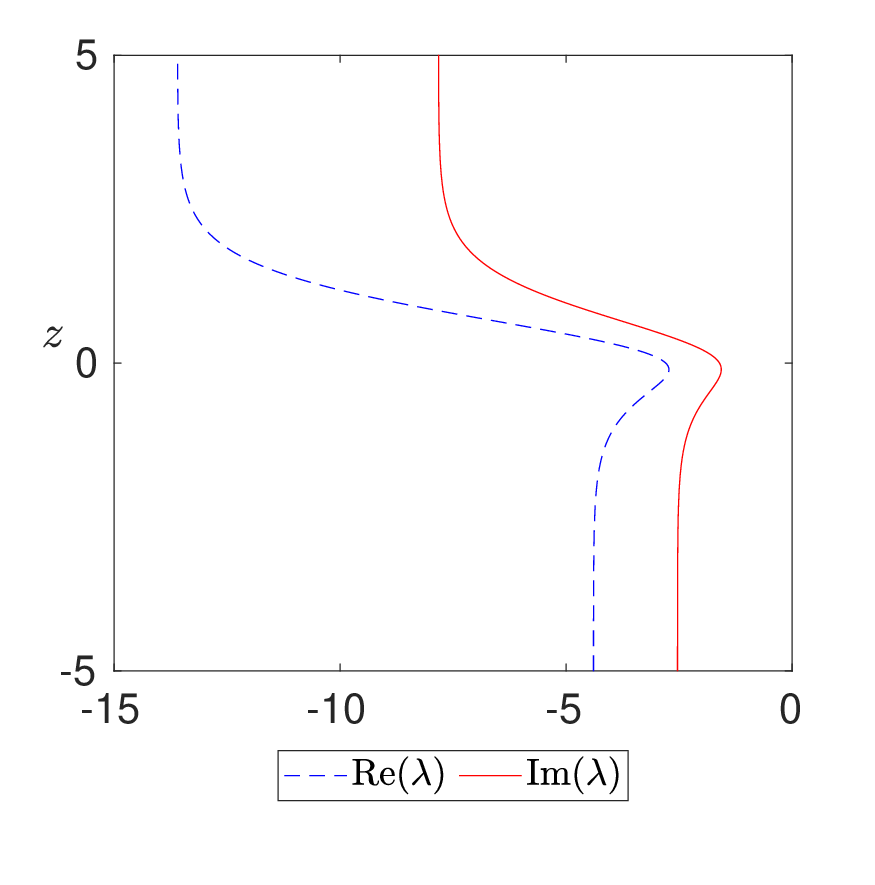} \hspace*{0.25cm}
\includegraphics[width=0.35\textwidth]{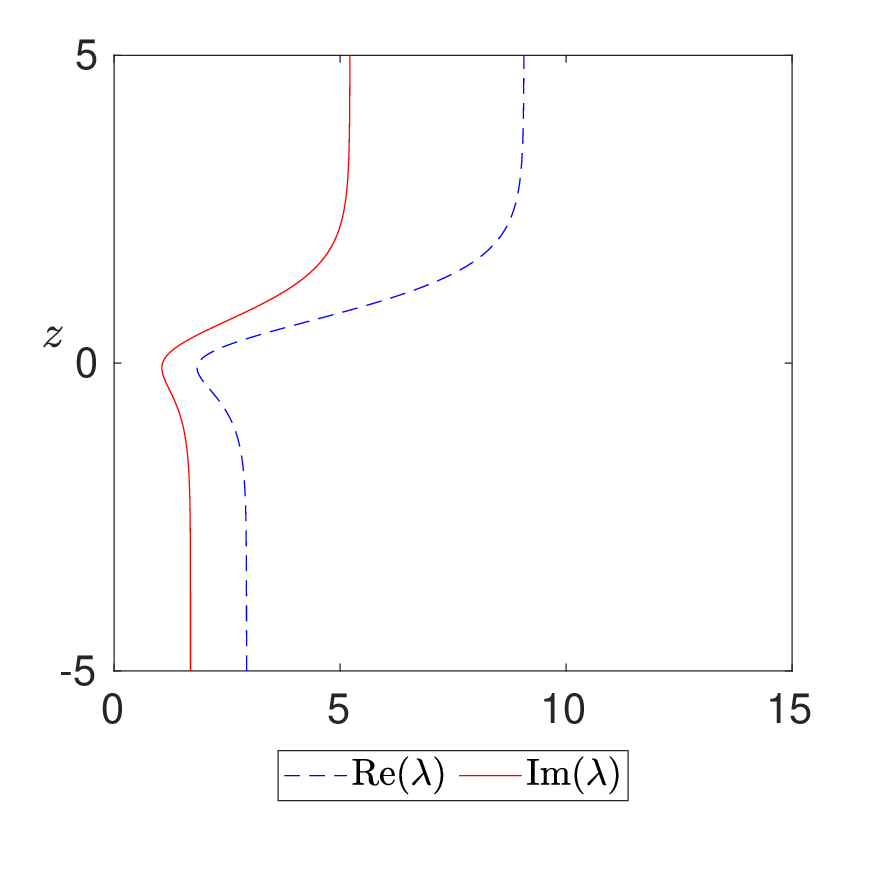} 
\includegraphics[width=0.35\textwidth]{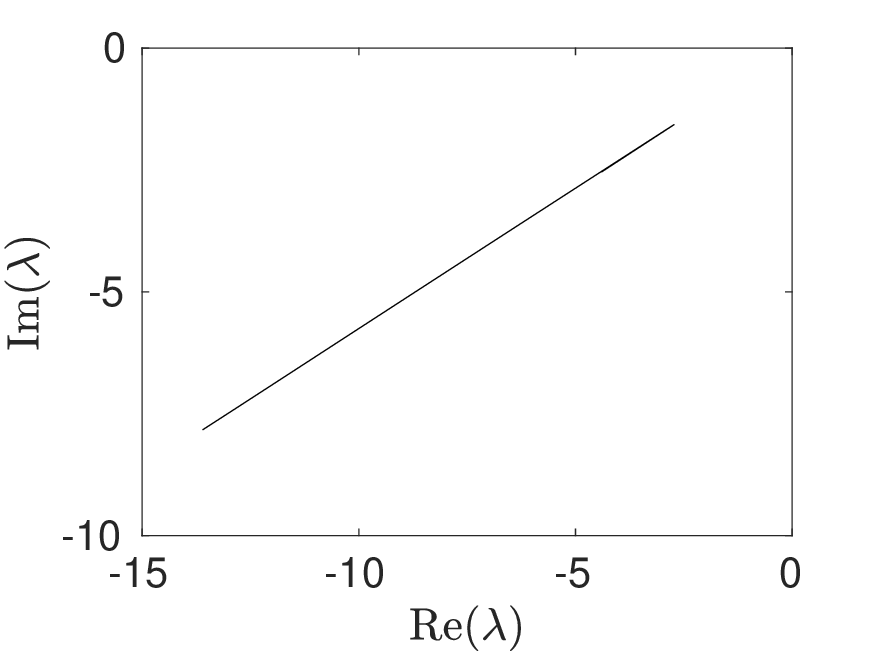} \hspace*{0.25cm}
\includegraphics[width=0.35\textwidth]{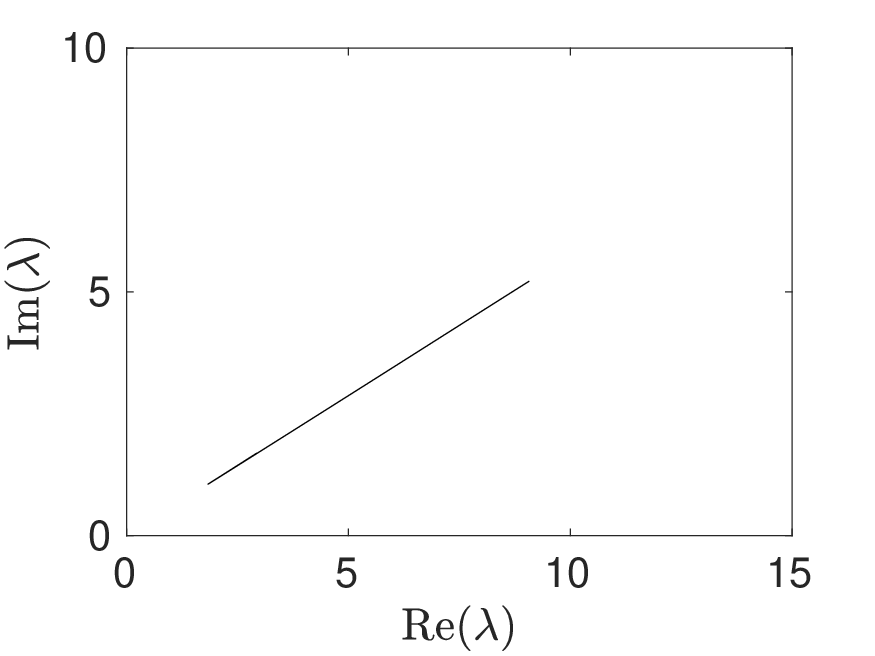} \vspace*{-0.5cm}
\end{center}
\caption{(Top panels) The real and imaginary parts of the eigenvalues $\lambda$ for the identical set parameters as in Figure~\ref{jwkbex}. (Bottom panels) The corresponding eigenvalue trajectories in the complex plane. (Left panels) Plots of the eigenvalues with negative real and imaginary values with $\varepsilon = 0.1$. (Right panels) Similar to the left panels but with Re$(\lambda) > 0$, Im$(\lambda) > 0$, and $\varepsilon = 0.15$.} \label{jwkbeigen}
\end{figure}

Consider the following well-known theorem on matrix exponential and fundamental matrix solutions from ODE theory~\cite{braun93,boyce12,edwards18,strang14}. Let $\boldsymbol{G}$ be an $n \times n$ matrix and let $\boldsymbol{\Phi}(z)$ be a fundamental matrix for the homogeneous linear system $\boldsymbol{x}' = \boldsymbol{G} \boldsymbol{x}$, then the unique solution of the initial (in this case, boundary) value problem (IVP or BVP)
\begin{equation}
\frac{d\boldsymbol{x}}{dz} = \boldsymbol{G} \boldsymbol{x}, \qquad \qquad \boldsymbol{x}(0) = \boldsymbol{x}_0,		\label{bvp}
\end{equation}
is given by
\begin{equation*}
\boldsymbol{x}(z) = e^{\boldsymbol{G}z} \boldsymbol{x}_0 = \boldsymbol{\Phi}(z) \left[\boldsymbol{\Phi}(0)\right]^{-1} \boldsymbol{x}_0.
\end{equation*}
Consider again the modified TG equation~\eqref{tago2} or~\eqref{tago3}. Defining $\tilde{v}' = \tilde{w}$, we can write it as a system of ODEs in the form of~\eqref{bvp}:
\begin{equation*}
\boldsymbol{x}' = \boldsymbol{G} \boldsymbol{x} \qquad \qquad \text{or} \qquad \qquad
\dot{\boldsymbol{x}} = \boldsymbol{G}_\varepsilon \boldsymbol{x}
\end{equation*}
where the prime and dot signs represent the derivative with respect to $z$ and $Z$, respectively and 
\begin{equation*}
\boldsymbol{x} = \begin{bmatrix*}[c]
\tilde{w} \\ \tilde{v}
\end{bmatrix*}, \qquad \qquad 
\boldsymbol{G} = \begin{bmatrix*}[c]
0 & 1 \\ -Q_2 & 0 
\end{bmatrix*},  \qquad \qquad \text{and} \qquad \qquad
\boldsymbol{G}_{\varepsilon} = \begin{bmatrix*}[c]
0 & 1 \\ -\frac{1}{\varepsilon^2} Q_\varepsilon & 0 
\end{bmatrix*}.
\end{equation*} 
The eigenvalues of $\boldsymbol{G}_\varepsilon$ are complex-valued functions that depends on the variable~$Z$:
{\small
\begin{equation*}
\lambda(Z) = \pm \sqrt{\frac{1}{2} \sqrt{\left(C - \frac{A}{\varepsilon^2}\right)^2 + \left(D - \frac{B}{\varepsilon^2}\right)^2} + C - \frac{A}{\varepsilon^2} } \pm i \sqrt{\frac{1}{2} \sqrt{\left(C - \frac{A}{\varepsilon^2}\right)^2 + \left(D - \frac{B}{\varepsilon^2}\right)^2} - \left(C - \frac{A}{\varepsilon^2} \right)}.
\end{equation*} 
}%
Figure~\ref{jwkbeigen} depicts the real and imaginary parts of eigenvalues $\lambda$ as well as their trajectories in the complex plane for two different values of~$\varepsilon$.

We need to seek the corresponding eigenvectors from these eigenvalues to obtain exact solutions using the fundamental matrix and matrix exponential, which turns out to be nontrivial. Instead, we will use the corresponding eigenfunctions from the WKB solution that we have obtained in the previous subsection to construct the physical optics approximation for the matrix exponential. Pursuing this course is particularly useful once we follow the path in solving the BVP numerically. One notable novel numerical integration technique combines the standard Runge-Kutta approach and the WKB method, known as the Runge-Kutta-Wentzel-Kramers-Brillouin (RKWKB) method~\cite{handley16,agocs20,bamber20}.

For the geometrical optics approximation of the WKB method, the complex-valued frequency can be approximated as 
\begin{align*}
\langle m_0 \rangle &= \frac{1}{z} \int_{0}^{z} m_0(\zeta) \, d\zeta, \qquad \qquad \qquad
\langle n_0 \rangle  = \frac{1}{z} \int_{0}^{z} n_0(\zeta) \, d\zeta \\
\langle M_0 \rangle &= \frac{1}{z} \int_{0}^{z} M_0(\zeta) \, d\zeta = \langle m_0 + i n_0 \rangle = \langle m_0 \rangle + i \langle n_0 \rangle \\
\langle \alpha \rangle &= \tan^{-1} \left(\frac{\langle m_0 \rangle}{\langle n_0 \rangle} \right).
\end{align*} 
The physical optics approximation provides a correction to the phase and amplitude of the solution. Using the fact that there are two linearly independent eigenfunction solutions, the matrix exponential approximation can be expressed as follows:
\begin{equation*}
e^{\boldsymbol{G}z} = \boldsymbol{\Phi}(z) \left[\boldsymbol{\Phi}(0)\right]^{-1}
= \begin{bmatrix*}[r]
	 \sqrt{\left|\frac{M_0(0)}{M_0(z)}\right|} \cosh \phi(z) & \frac{1}   {q(0)} \sqrt{\left|\frac{M_0(0)}{M_0(z)}\right|} \sinh \phi(z) \\
q(z) \sqrt{\left|\frac{M_0(0)}{M_0(z)}\right|} \sinh \phi(z) & \frac{q(z)}{q(0)} \sqrt{\left|\frac{M_0(0)}{M_0(z)}\right|} \cosh \phi(z)
\end{bmatrix*}
\end{equation*}
where the Wronskian $W(z) = \text{det} \, \boldsymbol{\Phi}(z) = -2q(z)/\sqrt{M_0(z)}$ and 
\begin{align*}
\phi(z) &= \langle M_0 \rangle z + \frac{1}{2} i \varepsilon \langle \alpha \rangle z \\
p(z) &= \left[-(m_0 m_0' + n_0 n_0') + i(m_0 n_0' - m_0'n_0)  \right] \\
q(z) &= M_0(z) + \frac{\varepsilon p(z)}{2 |M_0(z)|^2}.
\end{align*}
We observe that $\sqrt{\left|\frac{M_0(0)}{M_0(z)}\right|} \cosh \phi(z)$ and $\sqrt{\left|\frac{M_0(0)}{M_0(z)}\right|} \sinh \phi(z)$ play a role as two linearly independent eigenfunctions of the WKB solution. This is guaranteed from the fact that if the Wronskian starts from $W \neq 0$ at $z = 0$, then $W(z) \neq 0$ for all $z > 0$~\cite{strang14}.

\section{Behavior near turning point and critical level} \label{secbeha}

There are at least two possibilities where the WKB approximation becomes invalid, i.e., at a turning point or at a critical level. In this section, we will discuss the behavior of the perturbed vertical velocity profile near these points.

\subsection{Behavior near a turning point}

A \emph{turning point} is the point $z = z_0$ where $M_0$ vanishes, i.e., when it satisfies $N^2(z_0) = k^2 \left[U(z_0) - c\right] \left[U(z_0) - c - i k A_H(z_0) \right]$. At this incidence, the WKB solution~\eqref{phyopt} becomes singular. Let $f(z) = N^2 - k^2 u_1 u_2$, then its Taylor expansion about the turning point $z_0$ gives
\begin{equation*}
f(z) = f(z_0) + f'(z_0) (z - z_0) + \frac{1}{2} f''(z_0) (z - z_0)^2 + \dots .
\end{equation*}
Multiplying the TG equation~\eqref{tago2} with $u_1 u_2$, taking the first-order of Taylor-expansion of~$f$, and dividing again with $u_1 u_2$ after evaluating them at the turning point $z_0$, we obtain
\begin{equation*}
\frac{d^2 \tilde{w}}{dz^2} + \frac{f'(z_0)}{u_1(z_0) u_2(z_0)} (z - z_0) \tilde{w} = 0.
\end{equation*}
By defining a new variable $\zeta$:
\begin{equation*}
\zeta = \left(\frac{f'(z_0)}{u_1(z_0) u_2(z_0)} \right)^{1/3} (z_0 - z),
\end{equation*}
we arrive at the Airy ODE
\begin{equation}
\frac{d^2\tilde{w}}{d\zeta^2} - \zeta \tilde{w} = 0. \label{airyeqn}
\end{equation}

The Airy equation~\eqref{airyeqn} has two linearly independent solutions, called the Airy function of the first and second kinds, denoted by Ai$(\zeta)$ and Bi$(\zeta)$, respectively. Since Bi$(\zeta)$ blows up for $\zeta > 0$, this will not correspond to the physical setting of the stratified shear flows. We are more interested in Ai$(\zeta)$ because it decays for $\zeta > 0$ or for $z < z_0$. The function Ai$(\zeta)$ is defined by the following improper Riemann integral:
\begin{equation}
\text{Ai}(\zeta) = \frac{1}{\pi} \int_{0}^{\infty} \cos \left(\frac{t^3}{3} + \zeta t \right) \, dt.
\end{equation}
The convergence of this function can be shown by Dirichlet's test using the integration by parts~\cite{hall13}.

\subsection{Behavior near a critical level} \label{critic} 

A \emph{critical level} (or \emph{steering level}) is a height $z = z_c$, if it exists, at which the background state of the velocity flow $U(z)$ is equal to the real part of the horizontal phase speed $c$, i.e., $U(z_c) = c_r$. When this height is reached and $c \in \mathbb{R}$ ($c_i = 0$), the denominator of~$Q_{2}$ or~$Q_{\varepsilon}$ vanishes. Hence, the modified TG equation~\eqref{tago2} or~\eqref{tago3} becomes singular, and consequently, the WKB solution~\eqref{phyopt} vanishes too. Assume that $c_i$ is small, in particular, $|U''(z)| c_i/[U'(z)]^2 \ll 1$. For stable flow, the function $\tilde{w}(z)$ oscillate rapidly in $z$ near the critical level and the oscillations become infinitely rapid as the wave approaches it~\cite{haynes15}. Following an analysis presented in~\cite{booker67,nappo13}, we examine the behavior of the vertical velocity profile near the critical level.

Denoting $\zeta = z - z_c$, we proceed with Taylor-expanding the background wind speed $U$ as well as $U + ikA_H$ up to second-order terms
\begin{align}
U(z) - c \approx a_1 \zeta + \frac{1}{2} a_2 \zeta^2,& \quad \text{where} \; a_1 = U'(z_c) \; \text{and} \; a_2 = U''(z_c) \\
U(z) - i k A_H(z) - c \approx b_1 \zeta + \frac{1}{2} b_2 \zeta^2,& \quad \text{where} \; b_1 = a_1 - i \alpha_1\; \text{and} \; b_2 = a_2 - i \alpha_2.
\end{align}
Here, $\alpha_1 = k A_H'(z_c)$ and $\alpha_2 = k A_H''(z_c)$. Upon substitution to the modified TG equation~\eqref{tago2}, we obtain another TG equation with a regular singular point at $\zeta = 0$:
\begin{equation}
\tilde{w}'' + \left(\frac{Q_{22}}{\zeta^2} - \frac{Q_{21}}{\zeta} + Q_{20} \right) \tilde{w} = 0 \label{tago4}
\end{equation}
where
\begin{align}
Q_{22} &= \frac{N^2}{a_1 b_1} + \frac{1}{2} \left(\frac{a_1}{b_1} - 1 \right)\left(\frac{a_1}{b_1} - 2 \right) \\
Q_{21} &= \frac{1}{2} \left[ \frac{N^2}{a_1 b_1} \left(\frac{a_2}{a_1} + \frac{b_2}{b_1} \right) + \frac{b_2}{b_1} \left(\frac{a_1}{b_1} - 1 \right)\left(\frac{a_1}{b_1} - 2 \right) + \frac{a_2}{b_1} + 2 \frac{b_2}{b_1} \right] \\
Q_{20} &= \frac{1}{4} \left[N^2 \frac{a_2}{a_1^2} \cdot \frac{b_2}{b_1^2} + \frac{1}{2} \frac{b_2^2}{b_1^2} \left(\frac{a_1}{b_1} - 1 \right)\left(\frac{a_1}{b_1} - 2 \right) + \frac{a_2 b_2}{b_1^2} + 2 \frac{b_2^2}{b_1^2} \right] - k^2.
\end{align}

We seek a series solution near the critical level by employing the method of Frobenius by expanding $\tilde{w}$ as follows~\cite{braun93,boyce12,edwards18,frobenius73}:
\begin{equation}
\tilde{w}(z) = \tilde{w}(z_c + \zeta) = \sum_{n = 0}^{\infty} C_n \, \zeta^{n + \lambda}, \qquad  C_n \in \mathbb{C}. \label{frobe}
\end{equation}
Substituting~\eqref{frobe} to~\eqref{tago4} and ordering the terms according to the powers of $\zeta$ yields the indicial equation
$\lambda (\lambda - 1) + Q_{22} = 0$, with indicial roots or characteristic exponents at the singularity $\lambda$ at the regular singular point $\zeta = 0$ are given by
\begin{equation*}
\lambda = \frac{1}{2} \pm \rho_c e^{i\phi_c}, 
\end{equation*}
where
\begin{align*}
\rho_c &= \sqrt[4]{\text{Re}^2\left(\frac{1}{4} - Q_{22} \right) + \text{Im}^2\left(\frac{1}{4} - Q_{22} \right)} \\
\phi_c &= \frac{1}{2} \tan^{-1} \left[\frac{\text{Im}\left(\frac{1}{4} - Q_{22} \right)}{\text{Re}\left(\frac{1}{4} - Q_{22} \right)} \right] \\
\text{Re}\left(\frac{1}{4} - Q_{22} \right) &= \frac{1}{2} \frac{a_1^2}{|b_1|^4} \left(\alpha_1^2 - a_1^2 \right) + \frac{1}{|b_1|^2} \left(\frac{3}{2} a_1^2 - N^2 \right) - \frac{3}{4} \\
\text{Im}\left(\frac{1}{4} - Q_{22} \right) &= \frac{\alpha_1}{|b_1|^2} \left(\frac{3}{2} a_1 - \frac{N^2}{a_1} \right) - \alpha_1 \frac{a_1^3}{|b_1|^4}.
\end{align*}
We can express the solution of the modified TG equation~\eqref{tago4} as the following linear combination:
\begin{equation}
\tilde{w}(z) = \tilde{w}_1(z) \sum_{n = 0}^{\infty} A_n |z - z_c|^n + \tilde{w}_2 \sum_{n = 0}^{\infty} B_n |z - z_c|^n, \label{seriesfrobe}
\end{equation}	
where $\tilde{w}_1$ and $\tilde{w}_2$ are two linearly independent Frobenius solutions, given as follows:
\begin{align*}
\tilde{w}_1(z) &= |z - z_c|^{\frac{1}{2} + \rho_c \cos \phi_c} \cos \left(\rho_c \sin \phi_c \ln |z - z_c |\right)   \\ 
\tilde{w}_2(z) &= |z - z_c|^{\frac{1}{2} + \rho_c \cos \phi_c} \sin \left(\rho_c \sin \phi_c \ln |z - z_c |\right).
\end{align*}
The real-valued coefficients are given by $A_n = \text{Re}\left\{C_n \right\}$ and $B_n = \text{Im}\left\{C_n \right\}$, $n \in \mathbb{N}_0$, and the complex-valued coefficients $C_n$, $n \in \mathbb{N}_0$ satisfy the following recursive relationship:
\begin{equation}
\begin{aligned}
C_1 &= C_0 \frac{Q_{21}}{2 \lambda}, \qquad C_0 \in \mathbb{C} \; \text{is arbitrary}, \\
C_n &= \frac{C_{n - 1} Q_{21} - C_{n - 2} Q_{20}}{n (2 \lambda + n - 1)}, \qquad n \geq 2.
\end{aligned}		\label{recursive}
\end{equation}
The behavior of Frobenius solutions is rather nontrivial. Although both trigonometric terms are bounded, since they oscillate infinitely many times between their maximum and minimum values, they do not approach any single value as $z \rightarrow z_c$,~cf.~\cite{haynes15}. Additionally, we also need to ensure the positive power for the small term $|z - z_c|$ as the limit of $|z - z_c|$ for negative power as $z \rightarrow z_c$ does not exist.

In what follows, we will verify that the corresponding power series~\eqref{frobe} without the term~$\zeta^{\lambda}$ converges for all $|\zeta| < R$, where $R \geq 0$ is the radius of convergence of the term $(Q_{22} - Q_{21} \zeta + Q_{20} \zeta^2)$. In this particular case, $R = \infty$. Let $I(n + \lambda) = n (2 \lambda + n - 1)$ be a second-degree polynomial function in $n$ and $N_0 \in \mathbb{N}$ such that $I(n + \lambda) \neq 0$ for $n = N_0$, $N_0 + 1$, $N_0 + 2$, $\cdots$. The coefficients of the power series~\eqref{frobe} in the absence of $\zeta^{\lambda}$ satisfy the recursive relationship~\eqref{recursive}. 

Let us express the modified TG equation~\eqref{tago4} as $\tilde{w}'' + \zeta^{-2} \widetilde{Q}(\zeta) \tilde{w} = 0$, where
\begin{equation*}
\widetilde{Q}(\zeta) = \sum_{k = 0}^{\infty} \tilde{q}_k \zeta^k
\end{equation*}
with
\begin{equation*}
\tilde{q}_0 = Q_{22}, \qquad \tilde{q}_1 = - Q_{21}, \qquad \tilde{q}_2 = Q_{20}, \qquad \text{and} \qquad \tilde{q}_k = 0 \quad \text{for} \quad k = 3, 4, 5, \dots.
\end{equation*}
Note that this particular representation extends easily to the case where $\widetilde{Q}(\zeta)$ is any power series convergent for $|\zeta| < R$, where $R > 0$ is the corresponding radius of convergence. For any $\zeta \in (-R,R)$, choose $\Lambda_1 > 0$ and $\Lambda_2 > 0$ such that for $k \in \mathbb{N}$:
\begin{equation*}
|\zeta| < \Lambda_1 < \Lambda_2, \qquad \left|\tilde{q}_1 \Lambda_1 \right| < \Lambda_2, \qquad \text{and} \qquad \left| \tilde{q}_2 \Lambda_1^2 \right| < \Lambda_2.
\end{equation*}
Furthermore, since $\tilde{q}_k = 0$ for $k \geq 3$, we have
\begin{equation*}
\left| \tilde{q}_k \right| < \Lambda_2 \Lambda_1^{-k}, \qquad \text{for all} \; k \in \mathbb{N}.
\end{equation*}
We now consider a modified power series
\begin{equation}
\sum_{n = 0}^{\infty} \widetilde{C}_n \, \zeta^{n} 		\label{frobe2}
\end{equation}
with
\begin{equation*}
\widetilde{C}_0 = \left|C_0 \right|, \qquad \widetilde{C}_1 = \left|C_1 \right|, \qquad \text{and} \qquad
\widetilde{C}_n = \frac{1}{I(n + \lambda)} \sum_{j = 0}^{n - 1} \widetilde{C}_j \Lambda_2 \Lambda_1^{j-n} \qquad \text{for} \quad n \geq 2.
\end{equation*}
One could verify immediately that 
\begin{equation*}
\left|C_n \right| \, \left| \zeta^n \right| \leq \widetilde{C}_n \, \left|\zeta^n \right|, \qquad \text{for} \; n \in \mathbb{N}_0.
\end{equation*}
As a consequence, the convergence of the series~\eqref{frobe} without the factor $\zeta^{\lambda}$ can be confirmed by showing the convergence of the series~\eqref{frobe2} by means of the Ratio Test. Moreover, for $n > N_0$, we have
\begin{align*}
\widetilde{C}_{n + 1} &= \frac{1}{I(n + 1 + \lambda)} \sum_{j = 0}^{n} \widetilde{C}_j \Lambda_2 \Lambda_1^{j - n - 1} \\
&= \frac{1}{I(n + 1 + \lambda)} \left[ \left(\sum_{j = 0}^{n - 1} \widetilde{C}_j \Lambda_2 \Lambda_1^{j - n} \right) \Lambda_1^{-1} + \widetilde{C}_n \Lambda_2 \Lambda_1^{-1} \right] \\
&= \frac{I(n + \lambda) + \Lambda_2}{I(n + 1 + \lambda)} \, \widetilde{C}_n \Lambda_1^{-1}.
\end{align*}
Since $I$ is a second-degree polynomial in $n$, we can straightforwardly verify that 
\begin{equation*}
\lim\limits_{n \to \infty} \left| \frac{I(n + \lambda) + \Lambda_2}{I(n + 1 + \lambda)} \right| = 1.
\end{equation*}
Since $|\zeta| < \Lambda_1$, we obtain
\begin{equation*}
\lim\limits_{n \to \infty} \left|\frac{\widetilde{C}_{n+1} \ \zeta^{n + 1}}{\widetilde{C}_n \ \zeta^n} \right|
= \lim\limits_{n \rightarrow \infty} \frac{\widetilde{C}_{n + 1}}{\widetilde{C}_{n}} \ |\zeta|
= \lim\limits_{n \rightarrow \infty} \left|\frac{I(n + \lambda) + \Lambda_2}{I(n + 1 + \lambda)} \right| \ \frac{|\zeta|}{\Lambda_1} 
= 1.
\end{equation*}

Hence, the Frobenius series solution~\eqref{seriesfrobe} converges for $|\zeta| < R$. A similar proof for a more general second-order linear homogeneous ODEs and an exposition of different possible solutions obtained using the Frobenius method can be found in advanced texts on differential equations, e.g.,~\cite{agarwal09,birkhoff89,coddington89,howell19,teschl12}. Note that the Frobenius solution presented in~\cite{booker67,nappo13} only includes two terms of the lowest-order and discards the higher-order terms in the series. This suggests that those higher-order terms might not be essential and can be neglected from the physical point of view.

Finally, by employing the continuity equation~\eqref{eddyvoper}, applying the Liouville transformation~\eqref{liouv}, and restoring the normal mode, one may obtain the corresponding horizontal and vertical velocity fields.

\section{Conclusion} \label{conclusion} 
We have considered a modified TG equation describing a model for complex-valued vertical velocity perturbation in stratified shear flows with horizontal eddy coefficient of turbulent viscosity. Under an assumption of slowly varying background horizontal velocity and Brunt-V\"ais\"al\"a frequency, we implemented an asymptotic approach using the WKB method to obtain the geometrical and physical optics solutions for the modified TG equation. 

Furthermore, we also investigate the behavior of the WKB solution near a turning point and seek a series solution near a critical level by employing Frobenius' method. The former can be described by the Airy function while the latter exhibits an oscillating behavior but vanishes as the height approaches a critical level.

For the future work, it would be illuminating to discuss and analyze rigorously the relationship between the WKB method and a transformation at complex geometry. Although there exists an abundant literature on complex WKB method, e.g.,~\cite{voros83,maslov94,fedotov04}, most of them deals with the complex variable. In our case, however, the variable remains real-valued even though the potential-like function is complex-valued. Furthermore, we are interested in investigating the effect of vertical eddy coefficients of turbulent viscosity where its contribution is more essential in the study of atmospheric boundary layer in comparison to the horizontal one. Additionally, we would like to investigate the case when both horizontal and vertical eddy coefficients of turbulent diffusivity are present in the model equation.

\vspace{6pt} 

\funding{``This research received no external funding.''}

\acknowledgments{The author would like to thank Kenneth Howell for illuminating on the convergence proof of the Frobenius series, Vera Liu, Andy Chan, Ardhasena Sopaheluwakan, Francisco Marcelo Fernández, Lars Pesch, Mason Porter, Ramoshweu Solomon Lebelo, Shijun Liao, Sivasankaran Sivanandam, Charles Steele for fruitful discussions, and four anonymous referees for the suggestions in improving this article, whom two of them bringing the references~\cite{khani13,khani20,gubarev07,gubarev13} to our attention.}

\conflictsofinterest{``The authors declare no conflict of interest.''}

\abbreviations{The following abbreviations are used in this manuscript:\\

\noindent
\begin{tabular}{@{}ll}
ODE & ordinary differential equation \\
TG 	& Taylor-Goldstein \\
KH 	& Kevin-Helmholtz \\
WKB & Wentzel-Kramers-Brillouin
\end{tabular}}

\reftitle{References}

\end{document}